\documentclass[letterpaper]{article} % DO NOT CHANGE THIS
\usepackage{aaai25}  % DO NOT CHANGE THIS
\usepackage{times}  % DO NOT CHANGE THIS
\usepackage{helvet}  % DO NOT CHANGE THIS
\usepackage{courier}  % DO NOT CHANGE THIS
\usepackage[hyphens]{url}  % DO NOT CHANGE THIS
\usepackage{graphicx} % DO NOT CHANGE THIS
\usepackage{natbib}  % DO NOT CHANGE THIS AND DO NOT ADD ANY OPTIONS TO IT
\usepackage{caption} % DO NOT CHANGE THIS AND DO NOT ADD ANY OPTIONS TO IT
\usepackage{algorithm}
\usepackage{algpseudocode}
\usepackage{tabularx}
\usepackage{xcolor}
\usepackage{amssymb}
\usepackage{amsmath}
\usepackage{newfloat}
\usepackage{listings}
\DeclareCaptionStyle{ruled}{labelfont=normalfont,labelsep=colon,strut=off} % DO NOT CHANGE THIS
\floatstyle{ruled}
\newfloat{listing}{tb}{lst}{}
\floatname{listing}{Listing}
\title{Two-Player Zero-Sum Differential Games with One-Sided Information}
\author {
    Mukesh Ghimire\textsuperscript{\rm 1},
    Zhe Xu\textsuperscript{\rm 1},
    Yi Ren\textsuperscript{\rm 1}
}
\affiliations {
    \textsuperscript{\rm 1}Arizona State University\\
    \{mghimire, xzhe1, yiren\}@asu.edu
}
\usepackage{bibentry}
\begin{document}

\maketitle

\begin{abstract}
Unlike Poker where the action space $\mathcal{A}$ is discrete, differential games in the physical world often have continuous action spaces not amenable to discrete abstraction, rendering no-regret algorithms with $\mathcal{O}(|\mathcal{A}|)$ complexity not scalable. To address this challenge within the scope of two-player zero-sum (2p0s) games with one-sided information, we show that (1) a computational complexity independent of $|\mathcal{A}|$ can be achieved by exploiting the convexification property of incomplete-information games and the Isaacs' condition that commonly holds for dynamical systems, and that (2) the computation of the two equilibrium strategies can be decoupled under one-sidedness of information. Leveraging these insights, we develop an algorithm that successfully approximates the optimal strategy in a homing game. Code available in github\footnote{https://github.com/ghimiremukesh/cams/tree/workshop}.
\end{abstract}

%     \link{Code}{https://github.com/ghimiremukesh/cams/tree/workshop}
%     % \link{Datasets}{https://aaai.org/example/datasets}
%     % \link{Extended version}{https://aaai.org/example/extended-version}

\section{Introduction}
The strength of game solvers has grown rapidly in the last decade, beating elite-level human players in Chess~\citep{alphazero}, Go~\citep{alphago}, Poker~\citep{pluribus, rebel}, Diplomacy~\citep{cicero}, Stratego~\citep{stratego}, among others with increasing complexity. These successes motivated recent interests in solving differential games in continuous time and space, e.g., competitive sports~\citep{tacticai, ghimire24a}, where critical strategic plays should be executed precisely within the continuous action space and at specific moments in time (e.g., consider set piece scenarios in soccer).   
However, existing regret minimization algorithms, e.g., CFR+~\citep{tammelin2014solving} and its variants~\citep{burch2014solving, moravvcik2017deepstack,rebel,lanctot2009monte}, and last-iterate online learning algorithms, e.g., variants of follow the regularized leader (FTRL)~\citep{pmlr-v15-mcmahan11b, perolat2021poincare} and of mirror descent~\citep{sokota2022unified, cen2021fast, vieillard2020leverage}, are designed for discrete actions and have computational complexities increasing with respect to the size of the action space $\mathcal{A}$. Thus applying these algorithms to differential games would require either insightful action and time abstraction or enormous compute, neither of which are readily available.
% While most of the solution concepts can be applied here, the complexities grow exponentially due to continuous states and actions. Therefore, a scalable algorithm is imperative in solving differential games, especially with lack of information.

As a step towards addressing this challenge, our study focuses on games with one-sided information, which represent a variety of attack-defence scenarios: Both players have common knowledge about the finite set of $I$ possible payoff types and nature's distribution over these types $p_0$. At the beginning of the game, nature draws a type and informs Player 1 (P1) about the type but not P2. As the game progresses, the public belief about the chosen type is updated from $p_0$ based on the action sequence taken by P1 via the Bayes rule. P1's goal is to minimize the expected cost over $p_0$. This game is proved to have a value under Isaacs' condition~\citep{cardaliaguet2009numerical}. Due to the zero-sum nature, P1 may need to delay information release or manipulate P2's belief to take full advantage of information asymmetry, and P2's strategy is to optimize the worst-case payoff. 
Real-world examples of the game include man-on-man matchup in sports where the attacker has private information about which play is to be executed, and defense games where multiple potential targets are concerned.

\vspace{-0.07in}
\begin{figure}[!h]
    \begin{minipage}{0.56\linewidth}
        % \raggedright
        % \sloppy
\indent The two differences between our game and commonly studied imperfect-information extensive-form games (IIEFGs)~\citep{sandholm2010state, stratego, cicero} are that: (1) IIEFGs often have belief spaces (e.g., belief about opponent's cards in Poker) larger than their abstracted action spaces (e.g., betting categories in Poker), and (2) information asymmetry in our games is only one-sided. This paper investigates the potential computational advantages from exploiting these differences via the followi-
    \end{minipage}%
    \hfill
    \begin{minipage}{0.41\linewidth}
        \centering
        \vspace{-0.1in}
        \includegraphics[width=\linewidth]{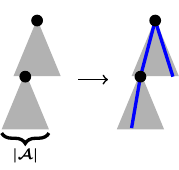}
        \vspace{-0.3in}
        \caption{\small SOTA algorithms like CFR require expanding over entire action space (left), whereas our algorithm only requires expanding over at most $I$ actions for P1 ($I+1$ for P2) at each decision node (right).}
        \label{fig:tree}
    \end{minipage}
    \vspace{-0.04in}
\end{figure}
\vspace{-0.14in}
 \noindent ng insights: (1) At any infostate, P1's (resp. P2's) behavioral strategy is $I$ (resp. $I+1$)-atomic and convexifies the primal (resp. dual) value with respect to the public belief (Fig.~\ref{fig:tree}). With this, we can reformulate the convex-concave minimax problem of size $\mathcal{O}(|\mathcal{A}|)$ at each infostate into a nonconvex-nonconcave problem of size $\mathcal{O}(I^2)$. When $I^2 \ll |\mathcal{A}|$, and in particular when $|\mathcal{A}| = \infty$, the latter becomes more efficient to solve in practice. (2) Due to the one-sidedness of information, the equilibrium behavioral strategies of P1 and P2 can be solved separately through primal and dual formulations of the game, in each of which the opponent plays pure best responses. This decoupling avoids recurrent learning dynamics between the pair of strategies without regularization~\citep{perolat2021poincare}.

To summarize, this work has two contributions: (1) familiarizing the broader AI community with the connections between computational game theory and differential game theory, and (2) providing the first algorithm with scalable convergence to the equilibrium of differential games with continuous action spaces and one-sided information.
\vspace{-0.1in}
\section{Related Work}
% add papers on online regret minimizations for related work
\paragraph{2p0s games with incomplete information.}
\cite{harsanyi1967games} introduced a Bayesian game framework to solve incomplete-information normal-form games by transforming the game into an imperfect-information one involving a chance mechanism. The seminal work of \cite{aumann1995repeated} extended this idea to repeated games and established the connection between value convexification and belief manipulation. Within the same framework, Blackwell's approachability theorem~\citep{blackwell1956analog} naturally becomes the theoretical support for the optimal strategy of the uninformed player (P2). Building on top of \cite{aumann1995repeated}, \cite{de1996repeated} introduced the concept of a dual game in which the behavioral strategy of the uninformed player becomes Markov. This concept later helped \cite{cardaliaguet2007differential,ghimire24a} to establish the value existence proof for 2p0s differential games with incomplete information. Unlike repeated games in which belief manipulation occurs only in the first round of the game, differential games may have multiple critical collocation points in the joint space of time, state, and public belief where belief manipulations are necessary to achieve Nash equilibrium, depending on the specifications of system dynamics, payoffs, and state constraints~\citep{ghimire24a}. For this reason, scalable value and strategy approximation for 2p0s differential games with incomplete information has not yet been achieved. 

% provided the theoretical foundation for characterizing how much information should be revealed and the gains that the revelation produces. 
% While repeated game captures the realistic decision-making scenario, it is still limited as it is assumed that the same normal-form game is repeated. 
\vspace{-0.05in}
\paragraph{Imperfect information extensive-form games.} IIEFGs represent the more general set of simultaneous or sequential multi-agent decision-making problems with finite horizons. Since any 2p0s IIEFG with finite action sets has a normal-form formulation, a unique Nash equilibrium always exists in the space of mixed strategies. Significant efforts have been taken to find equilibrium of large IIEFGs such as poker~\citep{koller1992complexity, billings2003approximating, gilpin2006finding, gilpin2007gradient, sandholm2010state, pluribus}, with a converging set of algorithms that are no-regret, average- or last-iterate converging, and with sublinear or linear convergence rates~\citep{zinkevich2007regret, abernethy2011blackwell, pmlr-v15-mcmahan11b, tammelin2014solving, johanson2012finding, lanctot2009monte, brown2019deep, brown2020combining, perolat2021poincare, sokota2022unified, stratego, sog} (see summary in Tab.~\ref{tab:complexity}). These algorithms all have computational complexities increasing with $|\mathcal{A}|$, provided that the equilibrium behavioral strategy lies in the interior of the simplex $\Delta(|\mathcal{A}|)$. Critically, this assumption does not hold for differential games equipped with the Isaacs' condition, in which case the equilibrium strategy is mostly pure along the game tree, and is atomic on $\mathcal{A}$ when mixed.
\begin{table}[h!]
    \centering
        \caption{Solver computational complexity with respect to action space $\mathcal{A}$ and equilibrium error $\varepsilon$}
        \vspace{-0.1in}
    \begin{tabularx}{\linewidth}{ X | p{0.24\linewidth} }
    \hline
         Algorithm & Complexity \\
         \hline
         CFR variants~\citep{zinkevich2007regret, lanctot2009monte, brown2019deep, tammelin2014solving,johanson2012finding} & $\mathcal{O}\left(\textcolor{red}{|\mathcal{A}|}\varepsilon^{-2}\right)$ to $\varepsilon$-Nash \\ \hline
         FTRL variants \& MMD ~\citep{pmlr-v15-mcmahan11b, perolat2021poincare, sokota2022unified} & $\mathcal{O}\left(\frac{\ln(\textcolor{red}{|\mathcal{A}|})}{\varepsilon}\ln\left(\frac{1}{\varepsilon}\right)\right)$ to $\varepsilon$-QRE \\
    \hline
    \end{tabularx}
    \label{tab:complexity}
    \vspace{-0.2in}
\end{table}
\paragraph{Descent-ascent algorithms for nonconvex-nonconcave minimax problems.} Existing developments in IIEFGs focused on convex-concave minimax problems due to the bilinear form of the expected payoff through the conversion of games to their normal forms. This paper, on the other hand, investigates the nonconvex-nonconcave minimax problems to be solved at every infostate when actions are considered continuous. To this end, we use the doubly smoothed gradient descent ascent method (DS-GDA) which has a worst-case complexity of $\mathcal{O}(\varepsilon^{-4})$
~\citep{zheng2023universal}.

\section{2p0s Differential Games w/ One-Sided Info.} \label{sec:diffgameintro}

\paragraph{Notations and preliminaries.} We use $\Delta(I)$ as the simplex in $\mathbb{R}^I$, $[K] :=\{1,...,K\}$ for $K \in \mathbb{Z}_+$, $a[i]$ as the $i$th element of vector $a$, $\partial V$ as the subgradient of function $V$, and $\langle \cdot,\cdot \rangle$ for vector product. Consider a time-invariant dynamical system that defines the evolution of the joint state $x \in \mathcal{X} \subseteq \mathbb{R}^{d_x}$ of P1 and P2 with control inputs $u \in \mathcal{U}$ and $v \in \mathcal{V}$, respectively:
\vspace{-0.05in}
\begin{equation}
    % \begin{aligned}
    %     \begin{cases}\dot{x}(t) &= f(x(t), u, v) \\
    %     % \quad u(t) \in \mathcal{U},\;v(t) \in \mathcal{V} \\
    %     x(t_0) &= x_0\end{cases}. 
    % \end{aligned}
    \dot{x}(t) = f(x(t), u, v).
    \label{eq:system}
    \vspace{-0.05in}
\end{equation}
% Without loss of generality, we consider $f$ to be time-invariant. 
The game starts at $t_0 \in [0, T]$ from some initial state $x(t_0) =x_0$. The initial belief $p_0 \in \Delta(I)$ is set to nature's distribution. P1 of type $i$ accumulates a running cost $l_i(u, v)$
% : \mathcal{U} \times \mathcal{V} \rightarrow \mathbb{R}$
during the game and receives a terminal cost $g_i(x(T))$, where $i \sim p_0$. The goal of P1 is to minimize the expected sum of the running and terminal costs, while P2 aims to maximize it. A behavioral strategy pair $(\eta, \zeta)$ is a Nash equilibrium (NE) of a zero-sum game if and only if 
\begin{equation}
    \inf_\eta \sup_\zeta \mathbb{E}_{\eta, \zeta, i \sim p_0} \int_0^T l_i dt + g_i = \sup_\zeta \inf_\eta \mathbb{E}_{\eta, \zeta, i \sim p_0} \int_0^T l_i dt + g_i,
    \label{eq:ne}
\end{equation}
and we call this common value the \textit{value of the game.} A NE is called \textit{pure} if the strategies $(\eta,\zeta)$ are deterministic, specifying a definite action for every decision point. It is called \textit{mixed} if the strategies are probabilistic, involving randomization over action spaces. When information is one-sided, $\eta = \{\eta_i\}^I$ since P1 prepares one strategy for each possible game type. 
% : \mathbb{R}^{d_x} \rightarrow \mathbb{R}$.
% $g_i (x(T))$. Optionally, the players also incur running costs $l(u(t), v(t))$, under which, P1 (and P2) would minimize (and maximize) the sum of the integral running costs and the final cost. 
% Before further developments, we list the key assumptions in this work:
We introduce the following assumptions under which mixed NE exists for Eq.~\eqref{eq:ne}~\citep{cardaliaguet2009numerical}:
\begin{enumerate}
    \item $\mathcal{U} \subseteq \mathbb{R}^{d_u}$ and $\mathcal{V} \subseteq \mathbb{R}^{d_v}$ are compact and finite-dimensional sets.
    \item $f: \mathcal{X} \times \mathcal{U} \times \mathcal{V} \rightarrow \mathcal{X}$ is bounded, continuous, and uniformly Lipschitz continuous with respect to $x$.
    \item $g_i: \mathcal{X} \rightarrow \mathbb{R}$ and $l_i: \mathcal{U} \times \mathcal{V} \rightarrow \mathbb{R}$ are Lipschitz continuous and bounded. 
    \item Isaacs' condition holds for the Hamiltonian $H: \mathcal{X} \times \mathbb{R}^{d_x} \rightarrow \mathbb{R}$:
    \begin{equation}\label{eq:isaacs}
        \begin{aligned}
            H (x, \xi) &:= \min_{u \in \mathcal{U}} \max_{v \in \mathcal{V}} f(x, u, v)^\top \xi - l_i(u, v) \\
            &= \max_{v \in \mathcal{V}} \min_{u \in \mathcal{U}}f(x, u, v)^\top \xi - l_i(u, v).
        \end{aligned}
        \vspace{-0.05in}
    \end{equation}
    \item Both players have full knowledge about $f$, $\{g_i\}_{i=1}^I$, $\{l_i\}_{i=1}^I$, $p_0$, and the NE of the game. Control inputs and states are fully observable and we assume perfect recall. 
\end{enumerate}
Critically, the Isaacs' condition ensures that 2p0s differential games with complete information have pure NE. 
% It is proved \citep{cardaliaguet2007differential, cardaliaguet2009numerical, ghimire24a} that under behavioral strategies $\eta = \{\eta_i\}^I$ and $\zeta$, the value of the game exists, and is of the form:
% \vspace{-0.1in}
% \begin{equation}\label{eq:primal_value}
% % \small
%     \begin{aligned}
%         V(t_0, x_0, p) &= \inf_{\eta_i \in \eta} \sup_{\zeta} \mathbb{E}_{\eta_i, \zeta} \left [\sum_{i=1}^Ip_i g_i\left(X_T^{t_0, x_0, \eta_i, \zeta}\right)\right. \\
%         &\hspace{0.5in}+\left. \int_{t_0}^T p_il_i(\eta_i(s), \zeta(s)) ds \right] \\
%         % &= \sup_{\zeta} \inf_{\eta_i \in \eta}\mathbb{E}_{\eta_i, \zeta} \left [\sum_{i=1}^Ip_i g_i\left(X_T^{t_0, x_0, \eta_i, \zeta}\right)\right. \\
%         % &\hspace{0.5in}+\left. \int_{t_0}^T p_il_i(\eta_i(s), \zeta(s)) ds \right].
%     \end{aligned}
% \end{equation}
\paragraph{Behavioral strategy of P1.}
A behavioral strategy prescribes distributions over the action space at every subgame $(t, x, p)$. In order to determine the strategy, it is necessary to first characterize the value function. From \citet{cardaliaguet2009numerical}, we obtain the following backward induction to approximate the value given a sufficiently fine time-discretization $\tau \rightarrow 0^+$:
\vspace{-0.08in}
\begin{equation}\label{eq:primal_backup}
% \small
    \begin{aligned}
        V_\tau(t, x, p) &= \text{Vex}_p\Big(\min_{u \in \mathcal{U}}\max_{v \in \mathcal{V}}V_\tau(t + \tau, x + \tau f(x, u, v), p)\\
        &\quad + \tau \mathbb{E}\;l(u, v)\Big);\; V_\tau(T, x, p) = \sum_i p_i g_i(x)\\
        \end{aligned}
    \vspace{-0.05in}
\end{equation}
where Vex is the convexification operator. 
The behavioral strategy of P1 is computed as follows: P1 first finds $\lambda = [\lambda^1, \dots, \lambda^I] \in \Delta(I)$ and $p^k \in \Delta(I)$ for $k \in [I]$ such that: 
\begin{equation} \label{eq:primal}
\small
    \begin{aligned}
        V_\tau(t, x, p) &= \sum_k \lambda_k \Big(\min_u \max_v \Big(V_\tau(t+\tau, x+\tau f(x, u, v), p)\\
        &\quad + \mathbb{E}\; l(u, v)\Big)\Big);\;\;  \sum_k \lambda_k p^k = p
    \end{aligned}
    \vspace{-0.01in}
\end{equation}
% Let $u^k$ be the non-revealing strategy of P1 corresponding to $p^k$. 
He then chooses $u^k$ with $ \Pr(u=u^k|i) = \lambda^k p^k[i]/p[i]$ if he is of type $i$ and updates the belief to $p^k$. This is famously known as the splitting mechanism in repeated game, and is a consequence of the ``Cav u'' theorem~\cite{aumann1995repeated, de1996repeated}.
\vspace{-0.05in}
\paragraph{Behavioral strategy of P2.}
For P2, the idea is to reformulate the game so that we can compute the value using P2's behavioral strategies and P1's pure best responses. This can be achieved by introducing the Fenchel conjugate $V^*$ of $V$:
\vspace{-0.07in}
\begin{equation}\label{eq:dual_value}
\small
    \begin{aligned}
        V^*(t_0,x_0,\hat{p}) &:=  \max_{p} p \cdot \hat{p} - V(t_0,x_0,p) \\
        &=  \inf_{\zeta} \sup_{\eta} \max_{i \in \{1, \dots, I\}} \left\{\hat{p}_i - \mathbb{E}_{\eta, \zeta} \left[ g_i\left(X_T^{t_0, x_0, \eta, \zeta}\right) \right.\right.\\
        &\quad +\left.\left. \int_{t_0}^T l_i(\eta(s), \zeta(s)) ds \right]\right\},
    \end{aligned}
\end{equation}
which describes a dual game with complete information in which P2's goal is to minimize some worst-case dual payoff. It is proved that P2's equilibrium in the dual game starting from some $(t_0,x_0,\hat{p})$ is also an equilibrium for the primal game if $\hat{p} \in \partial_p V(t_0,x_0,p)$ \cite{cardaliaguet2007differential}.

P2's strategy can be obtained through the dual game using a procedure similar to that of P1's: We first obtain the backward induction for the dual value:
\begin{equation}
\small
    \begin{aligned}
        V^*_\tau(T, x, \hat{p}) &= \max_i \{\hat{p}_i - g_i(x(T))\}\\
        V^*_\tau(t, x, \hat{p}) &= \text{Vex}_{\hat{p}}\Big(\min_{v \in \mathcal{V}}\max_{u \in \mathcal{U}}V^*_\tau(t + \tau, x + \tau f(x, u, v),\\
        &\hspace{1.6in}\hat{p} - \tau l(u, v))\Big),
    \end{aligned}
\end{equation}
where, $l = [l_1,...,l_I]^T$. Then at any $(t,x,\hat{p})$, P2 finds $\lambda = [\lambda^1, \dots, \lambda^{I+1}]$ and $\hat{p}^k \in \mathbb{R}^I$ for $k \in [I+1]$ such that:
\vspace{-0.1in}
\begin{equation}\label{eq:dual_backup}
% \small
    \begin{aligned}
        V^*_\tau(t, x, \hat{p}) &= \sum_k^{I+1} \lambda_k \Big(\min_v \max_u \Big(V^*_\tau(t + \tau, x + \tau f(x, u, v),\\
        &\hat{p} - \tau l(u, v))\Big);\; \sum_k^{I+1} \lambda_k \hat{p}^k = \hat{p}
    \end{aligned}
\end{equation}
% \vspace{-0.1in}
where $l = [l_1,...,l_I]^T$. P2's strategy is to compute the minimax solution $v^k$ corresponding to $\hat{p}^k$ and chooses $v = v^k$ with probability $\lambda^k$. 

\vspace{-0.15in}
\section{Methods}\label{sec:methods}
\paragraph{Reformulation of the primal and dual games.} 
% The goal of our algorithm is to efficiently compute the mixed strategy, which, in the case of the informed player is essentially is given by \eqref{eq:prob}. 
% We first summarize relationships among variables involved in \eqref{eq:primal}: 
To recap, at any $(t,x)$, P1 computes actions $u^k$ and their type-conditioned probabilities $\alpha_{ki} := \Pr(u = u^k|i)$ such that $\sum_{k=1}^I\alpha_{ki} = 1$ for $i \in [I]$.
% then we can represent the mixed strategy using a table of conditional probabilities. 
Then, $\lambda^k = \sum_{i=1}^I \alpha_{ki} p[i]$ and $p^k[i] = \alpha_{ki}p[i]/\lambda_k$ are both functions of $\alpha_{ki}$.
We can now reformulate \eqref{eq:primal} as follows:
% At any given time $t$, time-discretization $\tau$, state $x$, and belief $p$, P1 solves the following optimization problem:
\vspace{-0.05in}
\begin{equation}\label{eq:opt_prob} \tag{$\text{P}_1$}
\small
\begin{aligned}
    &\min_{\{u^k\}, \{\alpha_{ki}\}} \max_{\{v^k\}} \sum_{k=1}^{I}\lambda^k \left(V(t+\tau, x^k, p^k) + \tau \mathbb{E}_{i \sim p^k} [l_i(u^k, v^k)]\right)\\
    & \text{s.t. } u^k \in \mathcal{U}, \quad x^k = \text{ODE}(x, \tau, u^k, v^k; f), \quad v^k \in \mathcal{V}, \quad  \alpha_{ki} \in [0, 1], \; \\
    & \sum_{k=1}^I \alpha_{ki} = 1, \quad \lambda^k = \sum_{i=1}^I \alpha_{ki} p[i], \quad p^k[i] = \frac{\alpha_{ki}p[i]}{\lambda^k}, \quad \forall i, k \in [I].
\end{aligned}
\end{equation}
\ref{eq:opt_prob} is in general a nonconvex-nonconcave minimax problem of size $(\mathcal{O}(I(I+d_u)), \mathcal{O}(Id_v))$ that needs to be solved at all sampled infostates $(t, x, p)\in [0,T] \times \mathcal{X} \times \Delta(I)$.
% is formed by the resulting optimizers $\alpha_{ki}$ and $u^i$. When P1 announces his strategy, he picks the corresponding column of the matrix associated to his type. 
% The main advantage of CAMS is that it obviates the need for applying the Vex operator, by directly computing the convexified value. 
The resultant minimax objective is by definition the convexified value of the primal game. 
% This reformulation allows us to leverage the fact that at each decision node, P1 has at most $I$ actions to randomize from, which simplifies the game tree expansion as shown in Fig.~\ref{fig:tree}.
% Standard Gradient Ascent Descent (GDA) algorithms~\citep{jin2019minmax} along with its variants (SGDA~\citep{lin2020gradient}, DSGDA~\citep{zheng2023universal}) can be used to solve the minimax problem. 

% Note that CAMS is a ``dual-mode'' algorithm. It can be used for both value approximation as well as for strategy generation. In value approximation mode, we are interested in the optimal solution to the optimization problem, and in online play mode, we are interested in the optimizers.  
% \vspace{-0.2in}
% \begin{figure}[!h]
%     \centering
%     \includegraphics[width=0.45\linewidth]{CameraReady/LaTeX/figures/tree.pdf}
%     \vspace{-0.15in}
%     \caption{\small SOTA algorithms such as CFR require expanding over entire action space (left), whereas our algorithm only requires expanding over at most $I$ actions at each decision node (right).}
%     \label{fig:tree}
%     \vspace{-0.15in}
% \end{figure}
% explain 
P2, on the other hand, keeps track of the dual variable $\hat{p} \in \mathbb{R}^I$ instead of the public belief $p$ during the dual game and solves the following problem at all sampled infostates $(t, x, \hat{p})$: 
% Unlike P1, since P2 doesn't have any private information, she does not need to condition her actions. 
% P2's strategy consists of $\lambda_k$ and $v^k$ for $k \in [I+1]$. At any given time $t$, time-discretization $\tau$, state $x$, and $\hat{p}$, P2 solves the following optimization problem:
\begin{equation}\label{eq:opt_prob_dual} \tag{$\text{P}_2$}
\small
\begin{aligned}
    &\min_{\{v^k\}, \{\lambda^{k}\}, \{\hat{p}^k\}} \max_{\{u^k\}} \sum_{k=1}^{I+1}\lambda^k \left(V^*(t+\tau, x^k, \hat{p}^k - \tau l(u^k, v^k))\right)\\
    &\text{s.t. } u^k \in \mathcal{U}, \quad v^k \in \mathcal{V}, \quad x^k = \text{ODE}(x, \tau, u^k, v^k; f), \quad \lambda^{k} \in [0, 1],\\
    & \sum_{k=1}^{I+1} \lambda^{k}\hat{p}^k = \hat{p}, \quad \sum_{k=1}^{I+1}\lambda^k = 1, \quad k \in [I+1].\\
\end{aligned}
\end{equation}
\ref{eq:opt_prob_dual} is in general nonconvex-nonconcave of size $(\mathcal{O}(I(I+d_v), \mathcal{O}(Id_u ))$. 
\paragraph{Game solver.} We propose a continuous-action mixed-strategy (CAMS) solver for 2p0s differential games with one-sided information. Our algorithm performs Bellman backup through \ref{eq:opt_prob} (resp. \ref{eq:opt_prob_dual}) starting from the terminal condition in \eqref{eq:primal_backup} (resp. \eqref{eq:dual_backup}) at discretized time stamps $t \in \{T, T-\tau, ..., 0\}$ and $(x,p)$ (resp. $(x,\hat{p})$) uniformly sampled in $\mathcal{X} \times \Delta(I)$ (resp. $\mathcal{X} \times \mathbb{R}^I$). 
Specifically, at any $t$, with a value approximation model $\hat{V}_{t+1}: \mathcal{X} \times \Delta(I) \rightarrow \mathbb{R}$, we solve \ref{eq:opt_prob} using DS-GDA at $N$ collocation points $(x,p) \in \mathcal{X} \times \Delta(I)$ and collect a dataset $\mathcal{D}_t := \{(x^{(i)},p^{(i)},\tilde{V}^{(i)})\}_{i=1}^N$ where $\tilde{V}$ is the numerical approximation of the convexified value at $(t, x^{(i)},p^{(i)})$ for the minimax problem. Then we fit a model $\hat{V}_t(x,p)$ to $\mathcal{D}_t$ and go to $t-\tau$. 
Alg.~\ref{alg:cams} summarizes the solver for the primal game. The dual game solver is similarly defined.
% We note that our algorithm does not search along the game tree as in \citep{sog,rebel}. This is because value approximation over the entire $[0,T] \times \mathcal{X} \times \Delta(I)$ is necessary for long-term safety planning in risk-sensitive applications, e.g., when state constraints are imposed. We developed theory for solving such state-constrained games in \cite{ghimire24a}.    

% \begin{algorithm}
%     \caption{Continuous Action Mixed Strategy solver (CAMS)}
%     \label{alg:cams}
%     \begin{algorithmic}[1]
%         \Require time discretization $\tau$, terminal value $V(T,\cdot,\cdot)$, sample size $N$, minimax solver $\mathbb{O}$
%         \State Initialize value network $\{\hat{V}_t\}_{t=0}^{T-\tau}$, training dataset $\mathcal{D} \leftarrow \emptyset$
%         \State $\mathcal{S} \leftarrow$ sample $N$ states $(x, p) \in \mathcal{X} \times \Delta(I)$
%         \For{$t$ in $\{T-\tau,\dots,0\}$}
%             \For{each $(x, p)$ in $\mathcal{S}$}
%                 \State $\vartheta \leftarrow \mathbb{O}(t, x, p)$ \Comment{Solution to \ref{eq:opt_prob}}
%                 \State append $\{(t, x, p), \vartheta\}$ to $\mathcal{D}$
%             \EndFor
%             \State Fit $\hat{V}_{t}$ to $\mathcal{D}$
%         \EndFor
%     \end{algorithmic}
% \end{algorithm}
\begin{algorithm}
    \caption{Continuous Action Mixed Strategy Solver (CAMS)}
    \label{alg:cams}
    \begin{algorithmic}[1]
        \Require $\tau$, $V(T,\cdot,\cdot)$, $N$, minimax solver $\mathbb{O}$
        \State Initialize $\{\hat{V}_t\}_{t=0}^{T-\tau}$, $\mathcal{D} \leftarrow \emptyset$
        \State $\mathcal{S} \leftarrow$ sample $N$ states $(x, p) \in \mathcal{X} \times \Delta(I)$
        \For{$t \in \{T-\tau,\dots,0\}$}
            \For{$(x, p) \in \mathcal{S}$}
                \State Append $\{(t, x, p), \mathbb{O}(t, x, p)\}$ to $\mathcal{D}$
            \EndFor
            \State Fit $\hat{V}_t$ to $\mathcal{D}$
        \EndFor
    \end{algorithmic}
\end{algorithm}
\vspace{-0.25in}
\section{Empirical Validation}
\begin{figure}[!h]
    \begin{minipage}{0.55\linewidth}
        % \raggedright
      \paragraph{Comparison on 1- and 4-stage Hexner's Games.} We first use a normal-form Hexner's game with $\tau = T$ and a fixed initial state $x_0$ to demonstrate that IIEFG algorithms suffer from increasing costs along $|\mathcal{A}|$ while CAMS does not. We consider CFR+~\cite{tammelin2014solving}, MMD~\cite{sokota2022unified}, and a modified CFR-BR~\cite{johanson2012finding} (dubbed CFR-BR-Primal, where we only focus on
    \end{minipage}%
    \hfill
    \begin{minipage}{0.42\linewidth}
        \centering
        % \vspace{-0.05in}
        \includegraphics[width=\linewidth]{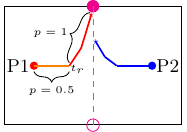}
        \caption{Hexner's game with a sample equilibrium trajectory. P1 starts to move to its target after $t_r$.}
        \label{fig:hexners_game}
    \end{minipage}
    \vspace{-0.08in}
\end{figure}

\begin{figure}[!h]
    \centering
    \vspace{-0.1in}
    \includegraphics[width=\linewidth]{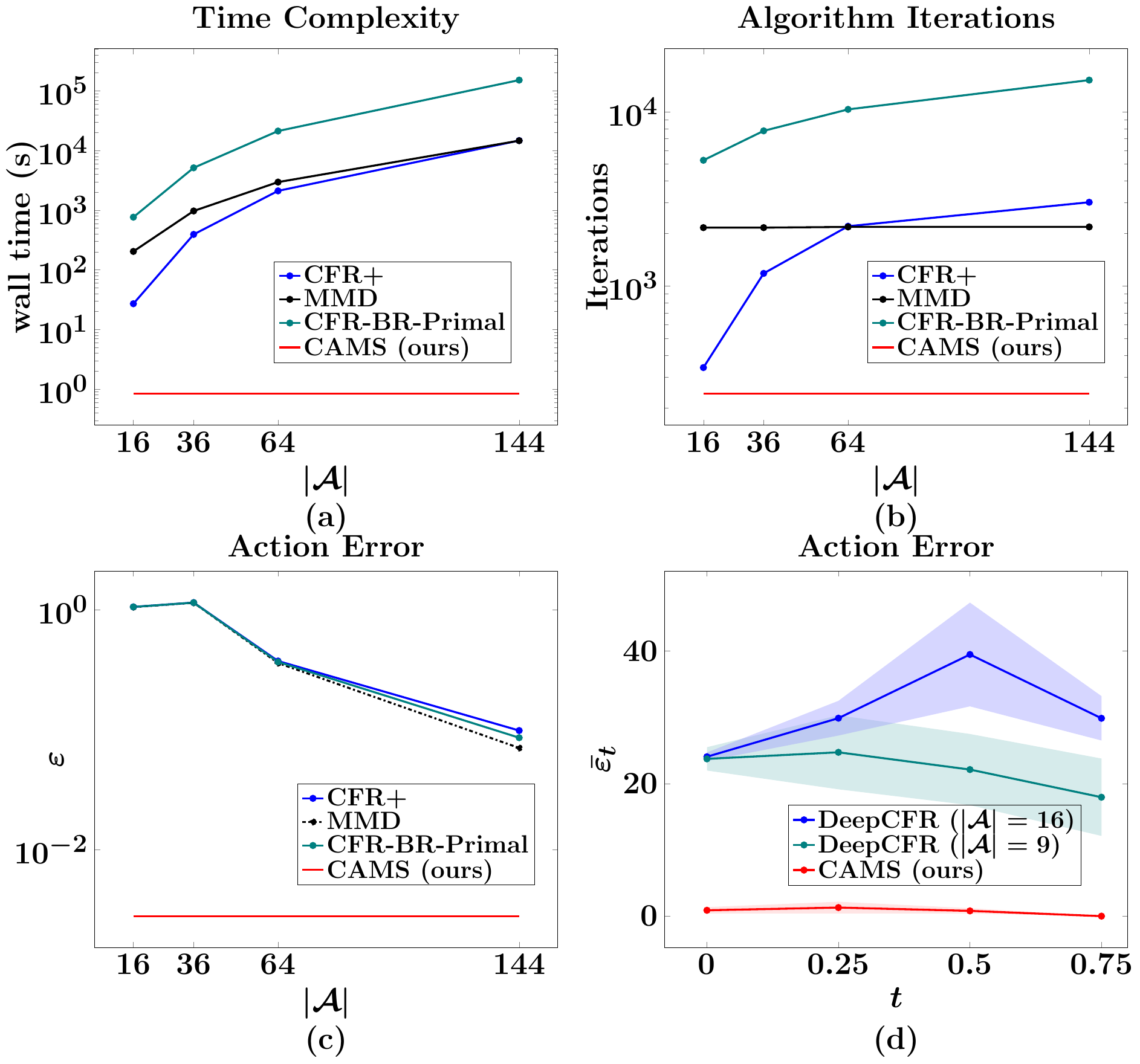}
    \vspace{-0.2in}
    \caption{Comparisons b/w CAMS and baseline algorithms.}
    \label{fig:benchmark}
\end{figure}

\begin{figure}[!h]
    \centering
    \includegraphics[width=\linewidth]{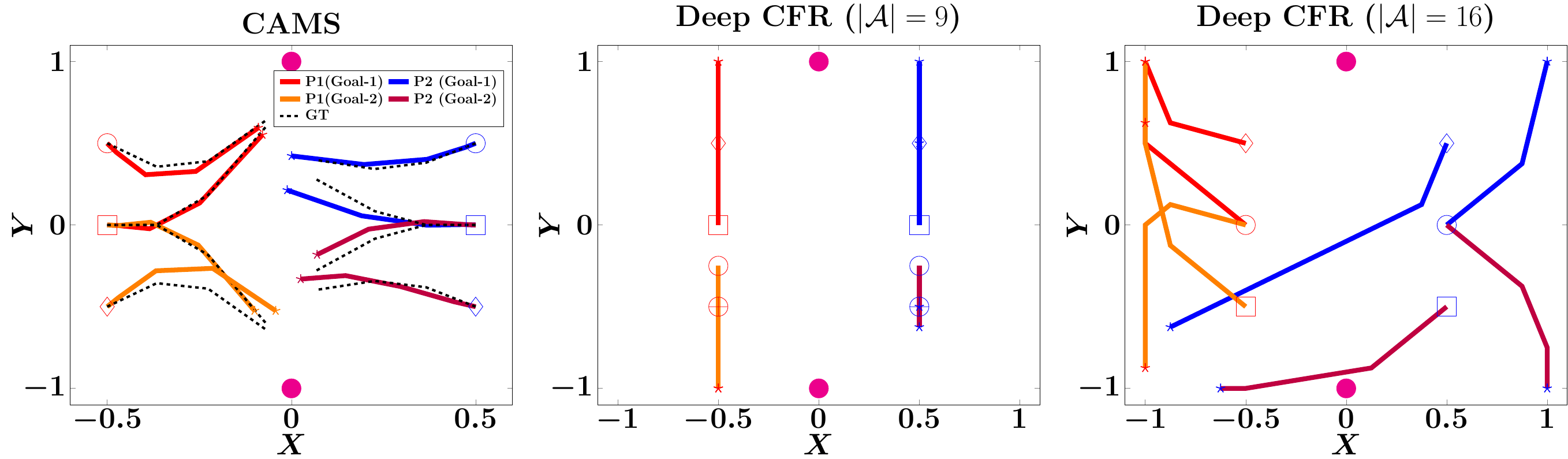}
    \caption{Trajectories using strategies from CAMS and DeepCFR. Markers indicate initial position.}
    \label{fig:traj}
    \vspace{-0.2in}
\end{figure}

We introduce Hexner's homing game~\citep{hexner1979differential} that has an analytical Nash equilibrium. We use variants of this game to compare CAMS with baselines (MMD, CFR+, and DeepCFR) on solution quality and computational cost. As shown in Fig.~\ref{fig:hexners_game}, it is a two-player game, in which P1's goal is to get closer to the target $\Theta$ unknown to P2, while keeping P2 away and minimizing running costs. The cost to P1 is the expected value of the total cost:
\vspace{-0.09in}
\begin{equation}
\small
\begin{aligned}
&J = \int_0^T (u^\top R_1 u - v^\top R_2 v)dt + [x_1(T) - \Theta]^\top K_1[x_1(T) - \Theta] \\
&\hspace{1.5in}- [x_2(T) - \Theta]^\top K_2[x_2(T) - \Theta],
\end{aligned}
\end{equation}
where $R_1\;, R_2 \succ 0$ and $K_1,\; K_2 \succeq 0$ are control and state-penalty matrices respectively. Due to the quadratic cost and decoupled dynamics, this game can be solved analytically as done in \citet{hexner1979differential}.
\vspace{-0.1in} 
\vspace{-0.1in}

   \noindent solving P1's optimal strategy) as baselines. Each player's state consists of 2D position and velocity. For baselines, we discretize the action sets $\mathcal{A}_1$ and $\mathcal{A}_2$ with sizes $\{16, 36, 64, 144\}$. All algorithms terminate when a threshold of NashConv (see \citet{lanctot2017unified} for definition) is met. For conciseness, we only consider solving P1's strategy and thus use P1's $\delta$ in NashConv. We then use DeepCFR as a baseline for a Hexner's game with 4 time-steps, where $T=1$ and $\tau = 0.25$. DeepCFRs were run for 1000 CFR iterations (resp. 100) with 10 (resp. 5) traversals for $|\mathcal{A}| = 9$ (resp. 16). We compare the computational cost and the expected action error $\varepsilon$ (and average action error at each time-step, $\bar{\varepsilon}_t$ for 4-stage game) from the ground-truth action of P1. Fig.~\ref{fig:benchmark} summarizes the comparisons. For the normal-form game, all baselines have complexities increasing with $\mathcal{A}$, while CAMS is invariant. In the 4-stage game, CAMS achieves significantly better strategies than DeepCFR, as visualized in Fig.~\ref{fig:traj}.
\vspace{-0.2in}
\vspace{-0.1in}
\section{Conclusion}
This work highlights the need for a scalable algorithm for solving incomplete-information differential games which are structurally similar to imperfect-information games such as poker. We demonstrated that SOTA IIEFG solvers are intractable when it comes to solving differential games. To the authors' best knowledge, this is the first method to provide tractable solution for incomplete-information differential games with continuous action spaces without problem-specific abstraction and discretization. 
\section{Acknowledgment}
This work is partially supported by NSF CNS 2304863, CNS 2339774, IIS 2332476, and ONR N00014-23-1-2505.
\bibliography{aaai25}

\begin{thebibliography}{37}
\providecommand{\natexlab}[1]{#1}

\bibitem[{Abernethy, Bartlett, and Hazan(2011)}]{abernethy2011blackwell}
Abernethy, J.; Bartlett, P.~L.; and Hazan, E. 2011.
\newblock Blackwell approachability and no-regret learning are equivalent.
\newblock In \emph{Proceedings of the 24th Annual Conference on Learning Theory}, 27--46. JMLR Workshop and Conference Proceedings.

\bibitem[{Aumann, Maschler, and Stearns(1995)}]{aumann1995repeated}
Aumann, R.~J.; Maschler, M.; and Stearns, R.~E. 1995.
\newblock \emph{Repeated games with incomplete information}.
\newblock MIT press.

\bibitem[{Billings et~al.(2003)Billings, Burch, Davidson, Holte, Schaeffer, Schauenberg, and Szafron}]{billings2003approximating}
Billings, D.; Burch, N.; Davidson, A.; Holte, R.; Schaeffer, J.; Schauenberg, T.; and Szafron, D. 2003.
\newblock Approximating game-theoretic optimal strategies for full-scale poker.
\newblock In \emph{IJCAI}, volume~3, 661.

\bibitem[{Blackwell(1956)}]{blackwell1956analog}
Blackwell, D. 1956.
\newblock An analog of the minimax theorem for vector payoffs.

\bibitem[{Brown et~al.(2020{\natexlab{a}})Brown, Bakhtin, Lerer, and Gong}]{rebel}
Brown, N.; Bakhtin, A.; Lerer, A.; and Gong, Q. 2020{\natexlab{a}}.
\newblock Combining deep reinforcement learning and search for imperfect-information games.
\newblock \emph{Advances in Neural Information Processing Systems}, 33: 17057--17069.

\bibitem[{Brown et~al.(2020{\natexlab{b}})Brown, Bakhtin, Lerer, and Gong}]{brown2020combining}
Brown, N.; Bakhtin, A.; Lerer, A.; and Gong, Q. 2020{\natexlab{b}}.
\newblock Combining deep reinforcement learning and search for imperfect-information games.
\newblock \emph{Advances in Neural Information Processing Systems}, 33: 17057--17069.

\bibitem[{Brown et~al.(2019)Brown, Lerer, Gross, and Sandholm}]{brown2019deep}
Brown, N.; Lerer, A.; Gross, S.; and Sandholm, T. 2019.
\newblock Deep counterfactual regret minimization.
\newblock In \emph{International conference on machine learning}, 793--802. PMLR.

\bibitem[{Brown and Sandholm(2019)}]{pluribus}
Brown, N.; and Sandholm, T. 2019.
\newblock Superhuman AI for multiplayer poker.
\newblock \emph{Science}, 365(6456): 885--890.

\bibitem[{Burch, Johanson, and Bowling(2014)}]{burch2014solving}
Burch, N.; Johanson, M.; and Bowling, M. 2014.
\newblock Solving imperfect information games using decomposition.
\newblock In \emph{Proceedings of the AAAI Conference on Artificial Intelligence}, volume~28.

\bibitem[{Cardaliaguet(2007)}]{cardaliaguet2007differential}
Cardaliaguet, P. 2007.
\newblock Differential games with asymmetric information.
\newblock \emph{SIAM journal on Control and Optimization}, 46(3): 816--838.

\bibitem[{Cardaliaguet(2009)}]{cardaliaguet2009numerical}
Cardaliaguet, P. 2009.
\newblock Numerical approximation and optimal strategies for differential games with lack of information on one side.
\newblock \emph{Advances in Dynamic Games and Their Applications: Analytical and Numerical Developments}, 1--18.

\bibitem[{Cen, Wei, and Chi(2021)}]{cen2021fast}
Cen, S.; Wei, Y.; and Chi, Y. 2021.
\newblock Fast policy extragradient methods for competitive games with entropy regularization.
\newblock \emph{Advances in Neural Information Processing Systems}, 34: 27952--27964.

\bibitem[{De~Meyer(1996)}]{de1996repeated}
De~Meyer, B. 1996.
\newblock Repeated games, duality and the central limit theorem.
\newblock \emph{Mathematics of Operations Research}, 21(1): 237--251.

\bibitem[{FAIR† et~al.(2022)FAIR†, Bakhtin, Brown, Dinan, Farina, Flaherty, Fried, Goff, Gray, Hu et~al.}]{cicero}
FAIR†, M. F. A. R. D.~T.; Bakhtin, A.; Brown, N.; Dinan, E.; Farina, G.; Flaherty, C.; Fried, D.; Goff, A.; Gray, J.; Hu, H.; et~al. 2022.
\newblock Human-level play in the game of Diplomacy by combining language models with strategic reasoning.
\newblock \emph{Science}, 378(6624): 1067--1074.

\bibitem[{Ghimire et~al.(2024)Ghimire, Zhang, Xu, and Ren}]{ghimire24a}
Ghimire, M.; Zhang, L.; Xu, Z.; and Ren, Y. 2024.
\newblock State-Constrained Zero-Sum Differential Games with One-Sided Information.
\newblock In Salakhutdinov, R.; Kolter, Z.; Heller, K.; Weller, A.; Oliver, N.; Scarlett, J.; and Berkenkamp, F., eds., \emph{Proceedings of the 41st International Conference on Machine Learning}, volume 235 of \emph{Proceedings of Machine Learning Research}, 15512--15539. PMLR.

\bibitem[{Gilpin et~al.(2007)Gilpin, Hoda, Pena, and Sandholm}]{gilpin2007gradient}
Gilpin, A.; Hoda, S.; Pena, J.; and Sandholm, T. 2007.
\newblock Gradient-based algorithms for finding Nash equilibria in extensive form games.
\newblock In \emph{Internet and Network Economics: Third International Workshop, WINE 2007, San Diego, CA, USA, December 12-14, 2007. Proceedings 3}, 57--69. Springer.

\bibitem[{Gilpin and Sandholm(2006)}]{gilpin2006finding}
Gilpin, A.; and Sandholm, T. 2006.
\newblock Finding equilibria in large sequential games of imperfect information.
\newblock In \emph{Proceedings of the 7th ACM conference on Electronic commerce}, 160--169.

\bibitem[{Harsanyi(1967)}]{harsanyi1967games}
Harsanyi, J.~C. 1967.
\newblock Games with incomplete information played by “Bayesian” players, I--III Part I. The basic model.
\newblock \emph{Management science}, 14(3): 159--182.

\bibitem[{Hexner(1979)}]{hexner1979differential}
Hexner, G. 1979.
\newblock A differential game of incomplete information.
\newblock \emph{Journal of Optimization Theory and Applications}, 28: 213--232.

\bibitem[{Johanson et~al.(2012)Johanson, Bard, Burch, and Bowling}]{johanson2012finding}
Johanson, M.; Bard, N.; Burch, N.; and Bowling, M. 2012.
\newblock Finding optimal abstract strategies in extensive-form games.
\newblock In \emph{Proceedings of the AAAI Conference on Artificial Intelligence}, volume~26, 1371--1379.

\bibitem[{Koller and Megiddo(1992)}]{koller1992complexity}
Koller, D.; and Megiddo, N. 1992.
\newblock The complexity of two-person zero-sum games in extensive form.
\newblock \emph{Games and economic behavior}, 4(4): 528--552.

\bibitem[{Lanctot et~al.(2009)Lanctot, Waugh, Zinkevich, and Bowling}]{lanctot2009monte}
Lanctot, M.; Waugh, K.; Zinkevich, M.; and Bowling, M. 2009.
\newblock Monte Carlo sampling for regret minimization in extensive games.
\newblock \emph{Advances in neural information processing systems}, 22.

\bibitem[{Lanctot et~al.(2017)Lanctot, Zambaldi, Gruslys, Lazaridou, Tuyls, P{\'e}rolat, Silver, and Graepel}]{lanctot2017unified}
Lanctot, M.; Zambaldi, V.; Gruslys, A.; Lazaridou, A.; Tuyls, K.; P{\'e}rolat, J.; Silver, D.; and Graepel, T. 2017.
\newblock A unified game-theoretic approach to multiagent reinforcement learning.
\newblock \emph{Advances in neural information processing systems}, 30.

\bibitem[{McMahan(2011)}]{pmlr-v15-mcmahan11b}
McMahan, B. 2011.
\newblock Follow-the-Regularized-Leader and Mirror Descent: Equivalence Theorems and L1 Regularization.
\newblock In Gordon, G.; Dunson, D.; and Dudík, M., eds., \emph{Proceedings of the Fourteenth International Conference on Artificial Intelligence and Statistics}, volume~15 of \emph{Proceedings of Machine Learning Research}, 525--533. Fort Lauderdale, FL, USA: PMLR.

\bibitem[{Morav{\v{c}}{\'\i}k et~al.(2017)Morav{\v{c}}{\'\i}k, Schmid, Burch, Lis{\`y}, Morrill, Bard, Davis, Waugh, Johanson, and Bowling}]{moravvcik2017deepstack}
Morav{\v{c}}{\'\i}k, M.; Schmid, M.; Burch, N.; Lis{\`y}, V.; Morrill, D.; Bard, N.; Davis, T.; Waugh, K.; Johanson, M.; and Bowling, M. 2017.
\newblock Deepstack: Expert-level artificial intelligence in heads-up no-limit poker.
\newblock \emph{Science}, 356(6337): 508--513.

\bibitem[{Perolat et~al.(2022)Perolat, De~Vylder, Hennes, Tarassov, Strub, de~Boer, Muller, Connor, Burch, Anthony et~al.}]{stratego}
Perolat, J.; De~Vylder, B.; Hennes, D.; Tarassov, E.; Strub, F.; de~Boer, V.; Muller, P.; Connor, J.~T.; Burch, N.; Anthony, T.; et~al. 2022.
\newblock Mastering the game of Stratego with model-free multiagent reinforcement learning.
\newblock \emph{Science}, 378(6623): 990--996.

\bibitem[{Perolat et~al.(2021)Perolat, Munos, Lespiau, Omidshafiei, Rowland, Ortega, Burch, Anthony, Balduzzi, De~Vylder et~al.}]{perolat2021poincare}
Perolat, J.; Munos, R.; Lespiau, J.-B.; Omidshafiei, S.; Rowland, M.; Ortega, P.; Burch, N.; Anthony, T.; Balduzzi, D.; De~Vylder, B.; et~al. 2021.
\newblock From poincar{\'e} recurrence to convergence in imperfect information games: Finding equilibrium via regularization.
\newblock In \emph{International Conference on Machine Learning}, 8525--8535. PMLR.

\bibitem[{Sandholm(2010)}]{sandholm2010state}
Sandholm, T. 2010.
\newblock The state of solving large incomplete-information games, and application to poker.
\newblock \emph{Ai Magazine}, 31(4): 13--32.

\bibitem[{Schmid et~al.(2023)Schmid, Morav{\v{c}}{\'\i}k, Burch, Kadlec, Davidson, Waugh, Bard, Timbers, Lanctot, Holland et~al.}]{sog}
Schmid, M.; Morav{\v{c}}{\'\i}k, M.; Burch, N.; Kadlec, R.; Davidson, J.; Waugh, K.; Bard, N.; Timbers, F.; Lanctot, M.; Holland, G.~Z.; et~al. 2023.
\newblock Student of Games: A unified learning algorithm for both perfect and imperfect information games.
\newblock \emph{Science Advances}, 9(46): eadg3256.

\bibitem[{Silver et~al.(2017{\natexlab{a}})Silver, Hubert, Schrittwieser, Antonoglou, Lai, Guez, Lanctot, Sifre, Kumaran, Graepel et~al.}]{alphazero}
Silver, D.; Hubert, T.; Schrittwieser, J.; Antonoglou, I.; Lai, M.; Guez, A.; Lanctot, M.; Sifre, L.; Kumaran, D.; Graepel, T.; et~al. 2017{\natexlab{a}}.
\newblock Mastering chess and shogi by self-play with a general reinforcement learning algorithm.
\newblock \emph{arXiv preprint arXiv:1712.01815}.

\bibitem[{Silver et~al.(2017{\natexlab{b}})Silver, Schrittwieser, Simonyan, Antonoglou, Huang, Guez, Hubert, Baker, Lai, Bolton et~al.}]{alphago}
Silver, D.; Schrittwieser, J.; Simonyan, K.; Antonoglou, I.; Huang, A.; Guez, A.; Hubert, T.; Baker, L.; Lai, M.; Bolton, A.; et~al. 2017{\natexlab{b}}.
\newblock Mastering the game of go without human knowledge.
\newblock \emph{nature}, 550(7676): 354--359.

\bibitem[{Sokota et~al.(2022)Sokota, D'Orazio, Kolter, Loizou, Lanctot, Mitliagkas, Brown, and Kroer}]{sokota2022unified}
Sokota, S.; D'Orazio, R.; Kolter, J.~Z.; Loizou, N.; Lanctot, M.; Mitliagkas, I.; Brown, N.; and Kroer, C. 2022.
\newblock A unified approach to reinforcement learning, quantal response equilibria, and two-player zero-sum games.
\newblock \emph{arXiv preprint arXiv:2206.05825}.

\bibitem[{Tammelin(2014)}]{tammelin2014solving}
Tammelin, O. 2014.
\newblock Solving large imperfect information games using CFR+.
\newblock \emph{arXiv preprint arXiv:1407.5042}.

\bibitem[{Vieillard et~al.(2020)Vieillard, Kozuno, Scherrer, Pietquin, Munos, and Geist}]{vieillard2020leverage}
Vieillard, N.; Kozuno, T.; Scherrer, B.; Pietquin, O.; Munos, R.; and Geist, M. 2020.
\newblock Leverage the average: an analysis of kl regularization in reinforcement learning.
\newblock \emph{Advances in Neural Information Processing Systems}, 33: 12163--12174.

\bibitem[{Wang et~al.(2024)Wang, Veli{\v{c}}kovi{\'c}, Hennes, Toma{\v{s}}ev, Prince, Kaisers, Bachrach, Elie, Wenliang, Piccinini et~al.}]{tacticai}
Wang, Z.; Veli{\v{c}}kovi{\'c}, P.; Hennes, D.; Toma{\v{s}}ev, N.; Prince, L.; Kaisers, M.; Bachrach, Y.; Elie, R.; Wenliang, L.~K.; Piccinini, F.; et~al. 2024.
\newblock TacticAI: an AI assistant for football tactics.
\newblock \emph{Nature communications}, 15(1): 1906.

\bibitem[{Zheng et~al.(2023)Zheng, Zhu, So, Blanchet, and Li}]{zheng2023universal}
Zheng, T.; Zhu, L.; So, A. M.-C.; Blanchet, J.; and Li, J. 2023.
\newblock Universal gradient descent ascent method for nonconvex-nonconcave minimax optimization.
\newblock \emph{Advances in Neural Information Processing Systems}, 36: 54075--54110.

\bibitem[{Zinkevich et~al.(2007)Zinkevich, Johanson, Bowling, and Piccione}]{zinkevich2007regret}
Zinkevich, M.; Johanson, M.; Bowling, M.; and Piccione, C. 2007.
\newblock Regret minimization in games with incomplete information.
\newblock \emph{Advances in neural information processing systems}, 20.

\end{thebibliography}

\end{document}